\documentclass{svmult}

\usepackage{makeidx}     
\usepackage{graphicx}    
\usepackage{epsf}
\usepackage{multicol}    

\makeindex             

\usepackage{amsmath, amssymb,latexsym,float,algorithm,algorithmic}
\usepackage{enumerate}

\newtheorem{thm}{Theorem}[section]

\newtheorem{prop}[thm]{Proposition}

\numberwithin{equation}{section}
\newcommand{\norm}[1]{\left\Vert#1\right\Vert}

\newcommand{\Real}{\mathbb R}
\newcommand{\Complex}{\mathbb C}

\newcommand{\EE}{\mathbb E}

\newcommand{\R}{\text{\fontshape{n}\selectfont I\kern-.42exR}}

\newcommand{\1}{\text{\fontshape{n}\selectfont 1\kern-.56exl}}

\begin{document}

\title*{Computational methods for the fermion determinant and the link between
overlap and domain wall fermions}
\titlerunning{Computational methods for the fermion determinant}
\author{Artan Bori\c{c}i\inst{}}
\institute{School of Physics\\
The University of Edinburgh\\
James Clerk Maxwell Building\\
Mayfield Road\\
Edinburgh EH9 3JZ\\
\hspace{1cm}\\
\texttt{borici@ph.ed.ac.uk}}
%
%
\maketitle

\vspace{1cm}
\abstract{This paper reviews the most popular
methods which are used in lattice QCD 
to compute the determinant of the lattice Dirac operator:
Gaussian integral representation and noisy methods. Both of them
lead naturally to matrix function problems. We review the most
recent development in Krylov subspace evaluation of matrix functions.
The second part of the paper reviews the formal relationship
and algebraic structure of domain wall and overlap fermions.
We review the multigrid algorithm to invert the overlap operator.
It is described here as a preconditioned Jacobi iteration where
the preconditioner is the Schur complement of a certain block of
the truncated overlap matrix.
}

\vspace{1cm}
\section{Lattice QCD}

Quantum Chromodynamics (QCD) is the quantum theory of interacting
quarks and gluons. It should explain the physics of strong force from
low to high energies. Due to asymptotic freedom of quarks at high
energies, it is possible to carry out perturbative calculations in QCD and
thus succeeding in explaining a range of phenomena. At low energies quarks
are confined within hadrons and the coupling between them is strong.
This requires non-perturbative calculations. The direct approach it is
known to be the lattice approach.
The lattice regularization of gauge theories was proposed
by \cite{Wi74}. It defines the theory in an Euclidean 4-dimensional
finite and regular lattice with periodic boundary conditions.
Such a theory is known to be Lattice QCD (LQCD).
The main task of LQCD is to compute the hadron spectrum and
compare it with experiment. But from the beginning it was
realized that numerical computation of the LQCD path integral is a
daunting task. Hence understanding the nuclear force has ever since
become a large-scale computational project.

In introducing the theory we will limit ourselves to
the smallest set of definitions that should allow a quick jump into the
computational tasks of LQCD.

A fermion field on a regular Euclidean lattice
$\Lambda$ is a Grassmann valued function
$\psi_{\mu,c}(x) \in G, ~x=\{ x_{\mu},\mu=1,\ldots,4 \} \in \Lambda$
which carries spin and colour indices $\mu = 1,\ldots,4, ~c = 1,2,3$.
Grassmann fields are anticommuting fields:
\begin{equation}
\psi_{\mu,c}(x) \psi_{\nu,b}(y) + \psi_{\nu,b}(y) \psi_{\mu,c}(x) = 0
\end{equation}
for $\psi_{\mu,c}(x), \psi_{\nu,b}(y) \in G$ and
$\mu,\nu = 1,\ldots,4, ~b,c = 1,2,3$. In the following we will denote
by $\psi(x) \in G_{12}$ the vector field with $12$ components corresponding to
Grassmann fields of different spin and colour index.

The first and second order differences are defined by the following expressions:

\begin{equation}
\begin{array}{l}
{\hat \partial}_{\mu} \psi(x) = \frac{1}{2 a} [\psi(x + a e_{\mu}) -
\psi(x - a e_{\mu})] \\
\\
{\hat \partial}_{\mu}^2 \psi(x) = \frac{1}{a^2} [\psi(x + a e_{\mu}) +
\psi(x - a e_{\mu}) - 2 \psi(x)]
\end{array}
\end{equation}
where $a$ and $e_{\mu}$ are the lattice spacing and the unit
lattice vector along the coordinate $\mu=1,\ldots,4$.

Let $U(x)_{\mu} \in \Complex^{3\times 3}$ be an unimodular unitary matrix,
an element of the $SU(3)$ Lie group in its fundamental
representation. It is a map onto $SU(3)$ colour group of the oriented link
connecting lattice sites $x$ and $x + a e_{\mu}$.
Physically it represents the gluonic field which mediates the quark
interactions represented by the Grassmann fields. A typical quark field interaction
on the lattice is given by the bilinear form:
\begin{equation}
{\bar \psi(x)} U(x)_{\mu} \psi(x + a e_{\mu})
\end{equation}
where $\bar \psi(x)$ is a second Grassmann field associated to $x \in \Lambda$.
Lattice covariant differences are defined by:

\begin{equation}
\begin{array}{l}
\nabla_{\mu} \psi(x) = \frac{1}{2 a} [U(x)_{\mu} \psi(x + a e_{\mu}) -
U^H(x - a e_{\mu})_{\mu} \psi(x - a e_{\mu})] \\
\\
\Delta_{\mu} \psi(x) = \frac{1}{a^2} [U(x)_{\mu} \psi(x + a e_{\mu}) +
U^H(x - a e_{\mu})_{\mu} \psi(x - a e_{\mu}) - 2 \psi(x)]
\end{array}
\end{equation}
where by $U^H(x)$ is denoted the Hermitian conjugation of the gauge
field $U(x)$, which acts on the colour components of the Grassmann fields.
The Wilson-Dirac operator is a matrix operator
$D_W(m_q,U) \in \Complex^{N\times N}$. It can be defined through $12\times 12$
block matrices $[D_W(m_q,U)](x,y) \in \Complex^{12\times 12}$ such that:
\begin{equation}\label{Wilson_op}
[D_W(m_q,U)\psi^q](x) = m_q \psi^q(x) + \sum_{\mu = 1}^4
~[\gamma_{\mu} \nabla_{\mu}\psi^q(x) - \frac{a}{2} \Delta_{\mu}\psi^q(x)]
\end{equation}
where $m_q$ is the bare quark mass with the index
$q = 1,\ldots,N_f$ denoting the quark flavour; $\psi^q(x)$ denotes the
Grassmann field corresponding to the quark flavour with mass $m_q$;
$\{ \gamma_{\mu} \in \Complex^{4\times 4},\mu=1,\ldots,5 \}$
is the set of
anti-commuting and Hermitian gamma-matrices of the Dirac-Clifford
algebra acting on the spin components of the Grassmann fields;
$N = 12 L_1 L_2 L_3 L_4$ is the total number of fermion fields
on a lattice with $L_1,L_2,L_3,L_4$ sites in each dimension. $D_W(m_q,U)$ is a
non-Hermitian operator. The Hermitian Wilson-Dirac operator is defined to be:
\begin{equation}
H_W(m_q,U) = \gamma_5 D_W(m_q,U)
\end{equation}
where the product by $\gamma_5$ should be understood as a product
acting on the spin subspace.

The fermion lattice action describing $N_f$ quark flavours is defined by:
\begin{equation}
S_f(U,\psi_1,\ldots,\psi_{N_f},{\bar \psi_1},\ldots,{\bar \psi_{N_f}}) =
\sum_{q=1}^{N_f}
\sum_{x,y \in \Lambda} {\bar \psi^q}(x) [D_W(m_q,U)](x,y) \psi^q(y)
\end{equation}
The gauge action which describes the dynamics of the gluon field
and its interaction to itself is given by:
\begin{equation}
S_g(U) = \frac{1}{g^2} \sum_{\mathbb{P}} \text{Tr} ~(\1 - U_{\mathbb{P}})
\end{equation}
where $\mathbb{P}$ denotes the oriented elementary square on the lattice
or the plaquette.
The sum in the right hand side is over all plaquettes with both orientations
and the trace is over the colour subspace.
$U_{\mathbb{P}}$ is a $SU(3)$ matrix defined on the plaquette $\mathbb{P}$
and $g$ is the bare coupling constant of the theory.

The basic computational task in lattice QCD is the evaluation of
the path integral:
\begin{equation}\label{Z_QCD}
Z_{QCD} = \int \sigma_H(U) \prod_{q=1}^{N_f} \sigma(\psi^q,{\bar \psi^q})
e^{-S_f(U,\psi_1,\ldots,\psi_{N_f},{\bar \psi_1},\ldots,{\bar \psi_{N_f}}) - S_g(U)}
\end{equation}
where $\sigma_H(U)$ and $\sigma(\psi^q,{\bar \psi^q})$ denote
the Haar and Grassmann measures for the $q$th
quark flavour respectively. The Haar measure is a
$SU(3)$ group character, whereas the Grassmann measure is defined using the rules of the
Berezin integration:
\begin{equation}
\int d\psi_{\mu,c}(x) = 0, ~~\int d\psi_{\mu,c}(x) \psi_{\mu,c}(x) = 1
\end{equation}
Since the fermionic action is a bilinear form on the Grassmann fields one gets:
\begin{equation}
Z_{QCD} = \int \sigma_H(U) \prod_{q=1}^{N_f} \det D_W(m_q,U) e^{-S_g(U)}
\end{equation}
Very often we take $N_f = 2$ for two degenerated `up' and `down' light quarks,
$m_u = m_d$.
In general, a path integral has $O(e^N)$ computational complexity which is
classified as an NP-hard computing problem \cite{WaWo96}.
But stochastic estimations of the
path integral can be done by $O(N^{\alpha})$ complexity with $\alpha \geq 1$.
This is indeed the case for the Monte Carlo methods that are used extensively
in lattice QCD, a topic which is reviewed in
this volume by Mike Peardon \cite{Pe03}.

It is clear now that the bottle-neck of any computation in lattice QCD is the
complexity of the fermion determinant evaluation. A very often made
approximation is to ignore the determinant altogether. Physically this
corresponds to a QCD vacuum without quarks, an approximation which
gives errors of the order 10\% in the computed mass spectrum. This
is called the valence or quenched approximation which requires modest computing
resources compared to the true theory. To answer the critical
question whether QCD is the theory of quarks and gluons it is thus
necessary to include the determinant in the path integral evaluation.

Direct methods to compute the determinant of a large
and sparse matrix are very expensive and even not adequate for this class
of matrices. The complexity of $LU$ decomposition is $O(N^3)$ and it is not
feasible for matrices with $N = 1920000$ which is the case for a lattice with
$20$ sites across each dimension. Even $O(N^2)$ methods are still very expensive.
Only $O(N)$ methods are feasible for the present computing power
for such a large problem.

\section{Gaussian integral representation: pseudofermions}
\label{sec:2}

The determinant of a positive definite matrix, which can be diagonalized has a
Gaussian integral representation. We assume here that we are dealing with a matrix
$A \in \Complex^{N\times N}$ which is Hermitian and positive definite. For example,
$A = H_W(m_q,U)^2$. It is easy to show that:
\begin{equation}
\det A = \int \prod_{i=1}^N \frac{d\text{Re}(\phi_i) d\text{Im}(\phi_i)}{\pi}
~~e^{-\phi^H A^{-1} \phi}
\end{equation}
The vector field
$\phi(x) \in \Complex^{12}, x \in \Lambda$ that went under the name pseudofermion
field \cite{WePe81}, has the structure of a fermion field but its components are
complex numbers (as opposed to Grassmann numbers for a fermion field).

Pseudofermions have obvious advantages to work with. One can use iterative
algorithms to invert $A$ which are well suited for large and sparse problems.
The added complexity of an extended variable space of the integrand
can be handled easily by Monte Carlo methods.

However, if $A$ is
ill-conditioned then any $O(N^{\alpha})$ Monte Carlo algorithm, which is used
for path integral evaluations is bound to produce small changes in the gauge field.
(Of course, an $O(e^N)$ algorithm would allow changes of any size!)
Thus, to produce the next statistically independent 
gauge field one has to perform a large number of matrix inversions which
grows proportionally with the condition number.
Unfortunately, this is the case in lattice QCD since the unquenching
effects in hadron spectrum are expected to come form light quarks
which in turn make the Wilson-Dirac matrix nearly singular.

The situation can be improved if one uses fast inversion
algorithms. This was the hope in the early '90 when state of the art
solvers were probed and researched for lattice QCD \cite{BoFo94,Fr_et_al94}.
Although revolutionary for the lattice community of that time,
these methods alone could not improve
significantly the above picture.

Nonetheless, pseudofermions remain
the state of the art representation of the fermion determinant.

\section{Noisy methods}

Another approach that was introduced later is the noisy estimation of the
fermion determinant \cite{Bai_et_al,Th_et_al98,Cahill_et_al,Borici03_comp}.
It is based on the identity:
\begin{equation}
\det A = e^{Tr \log A}
\end{equation}
and the noisy estimation of the trace of the natural logarithm of $A$.

Let $Z_j \in \{+1,-1\}, j = 1,\ldots,N$ be independent
and identically distributed random variables with probabilities:
\begin{equation}
\text{Prob}(Z_j = 1) = \text{Prob}(Z_j = -1) = \frac{1}{2}, ~~~j = 1, \ldots,N
\end{equation}
Then for the expectation values we get:
\begin{equation}
\EE(Z_j) = 0, ~~~\EE(Z_jZ_k) = \delta_{jk}, ~~~j,k = 1, \ldots, N
\end{equation}
and the following result holds:
\begin{prop}
Let $X$ be a random variable defined by:
\begin{equation}
X = Z^T \log A Z, ~~Z^T = (Z_1,Z_2,\ldots,Z_N)
\end{equation}
Then its expectation $\mu$ and variance $\sigma^2$ are given by:
\begin{equation}
\mu = \EE(X) = Tr \log A,
~~~\sigma^2 = \EE[(X-\mu)^2] = 2 \sum_{j\neq k} [Re(\log A)_{jk}]^2
\end{equation}
\end{prop}

To evaluate the matrix logarithm one can use the methods described in
\cite{Bai_et_al,Th_et_al98,Cahill_et_al,Borici03_comp}.
These methods have similar complexity
with the inversion algorithms and are subject of the next section.

However, noisy methods give a biased estimation of the determinant.
This bias can be reduced by reducing the variance of the estimation.
A straightforward way to do this is to take a sample of estimations
$X_1,\ldots,X_p$ and to take as estimator their arithmetic mean.

\cite{Th_et_al98} subtract traceless matrices which reduce the error
on the determinant from $559\%$ to $17\%$. \cite{Go_qcdna_03} proposes
a promising control variate technique which can be found in this volume.

Another idea is to suppress or `freeze´ large eigenvalues of the fermion
determinant. They are known to be artifacts of a discretized differential
operator. This formulation reduces by an order of magnitude unphysical
fluctuations induced by lattice gauge fields \cite{Borici03_new_op}.

A more radical approach is to remove the bias altogether. The idea is
to get a noisy estimator of $Tr \log A$ by choosing a certain order
statistic $X_{(k)} \in \{X_{(1)} \leq X_{(2)} \ldots \leq X_{(p)}\}$
such that the determinant estimation is unbiased \cite{Borici_lat03}.
More one this subject can be found in this volume from the same author
\cite{Borici_qcdna_03}.

\section{Evaluation of bilinear forms of matrix functions}

We describe here a Lanczos method for evaluation of bilinear forms of the type:
\begin{equation}\label{b_forms}
{\mathcal F}(b,A) = b^T f(A) b
\end{equation}
where $b \in \Real^N$ is a random vector and
$f(s)$ is a real and smooth function of $s \in \Real_+$.

The Lanczos method described here is similar to the method
of \cite{Bai_et_al}. Its viability for lattice QCD computations
has been demonstrated in the recent work of \cite{Cahill_et_al}.
\cite{Bai_et_al} derive their method using quadrature rules and Lanczos
polynomials. Here, we give an alternative derivation which
is based on the approach of \cite{Borici_over,Borici_WUP,Borici_isqr}.
The Lanczos method enters the derivation as an algorithm
for solving linear systems of the form:
\begin{equation}\label{lin_sys}
A x = b, ~~~~~~x \in \Complex^N
\end{equation}

\paragraph{Lanczos algorithm}

$n$ steps of the Lanczos algorithm \cite{Lanczos} on the
pair $(A,b)$ are given by Algorithm \ref{Lanczos_algor}.
\begin{algorithm}[htp]
\caption{The Lanczos algorithm}
\label{Lanczos_algor}
\begin{algorithmic}
\STATE Set $\beta_0 = 0, ~q_0 = o, ~q_1 = b / ||b||^2$
\FOR{$~i = 1, \ldots n$}
    \STATE $v = A q_i$
    \STATE $\alpha_i = q_i^{\dag} v$
    \STATE $v := v - q_i \alpha_i - q_{i-1} \beta_{i-1}$
    \STATE $\beta_i = ||v||_2$
    \STATE $q_{i+1} = v / \beta_i$
\ENDFOR
\end{algorithmic}
\end{algorithm}

The Lanczos vectors $q_1, \ldots, q_n \in \Complex^N$
can be compactly denoted by the matrix $Q_n = [q_1, \ldots, q_n]$.
They are a basis of the Krylov subspace
$\mathcal{K}_n = \text{span}\{b,Ab,\ldots,A^{n-1}b\}$.
It can be shown that the following identity holds:
\begin{equation}\label{AQ_QT}
A Q_n = Q_n T_n + \beta_n q_{n+1} e_n^T, ~~~~~q_1 = b/||b||_2
\end{equation}
$e_n$ is the last column of the identity matrix ${\1}_n \in \Real^{n\times n}$
and $T_n$ is the tridiagonal and symmetric matrix given by:
\begin{equation}\label{T_n}
T_n =
\begin{pmatrix} \alpha_1 & \beta_1  &             &             \\
                \beta_1  & \alpha_2 & \ddots      &             \\
                         & \ddots   & \ddots      & \beta_{n-1} \\
                         &          & \beta_{n-1} & \alpha_n    \\
\end{pmatrix}
\end{equation}
The matrix (\ref{T_n}) is often referred to as the Lanczos matrix.
Its eigenvalues, the so called Ritz values, tend to approximate the
extreme eigenvalues of the original matrix $A$ as $n$ increases.

To solve the linear system (\ref{lin_sys}) we seek an approximate solution
$x_n \in \mathcal{K}_n$ as a linear combination of the
Lanczos vectors:
\begin{equation}\label{x_n}
x_n = Q_n y_n, ~~~~~y_n \in \Complex^n
\end{equation}
and project the linear system (\ref{lin_sys}) on to the
Krylov subspace $\mathcal{K}_n$:
\begin{equation*}
Q_n^{\dag} A Q_n y_n = Q_n^{\dag} b = Q_n^{\dag} q_1 ||b||_2
\end{equation*}
Using (\ref{AQ_QT}) and the orthonormality of Lanczos vectors, we obtain:
\begin{equation*}
T_n y_n = e_1 ||b||_2
\end{equation*}
where $e_1$ is the first column of the identity matrix ${\1}_n$.
By substituting $y_n$ into (\ref{x_n}) one obtains the approximate
solution:
\begin{equation}\label{x_sol}
x_n = Q_n T_n^{-1} e_1 ||b||_2
\end{equation}

The algorithm of \cite{Th_et_al98} is based on
the Pad\'e approximation of the smooth and bounded
function $f(.)$ in an interval \cite{Graves-Morris}.
Without loss of generality one can assume a diagonal
Pad\'e approximation in the interval $s \in (0,1)$.
It can be expressed as a partial fraction expansion. Therefore,
one can write:
\begin{equation}
f(s) \approx \sum_{k=1}^m \frac{c_k}{s + d_k}
\end{equation}
with $c_k \in \Real, d_k \geq 0, k = 1, \ldots, m$.
Since the approximation error $O(s^{2m+1})$ can be
made small enough as $m$ increases, it can be assumed
that the right hand side converges to the left hand side
as the number of partial fractions becomes large enough.
For the bilinear form we obtain:
\begin{equation}\label{partial_frac}
{\mathcal F}(b,A) \approx \sum_{k=1}^m b^T \frac{c_k}{A + d_k \1} b
\end{equation}

Having the
partial fraction coefficients one can use a multi-shift iterative
solver of \cite{Freund} to evaluate the right hand side (\ref{partial_frac}).
To see how this works, we solve the shifted linear system:
\begin{equation*}
(A + d_k \1) x^k = b
\end{equation*}
using the same Krylov subspace $\mathcal{K}_n$. A closer inspection
of the Lanczos algorithm, Algorithm \ref{Lanczos_algor} suggests that
in the presence of the shift $d_k$ we get:
\begin{equation*}
\alpha_i^k = \alpha_i + d_k
\end{equation*}
while the rest of the algorithm remains the same. This is the so called
shift-invariance of the Lanczos algorithm. From this property and by
repeating the same arguments which led to (\ref{x_sol}) we get:
\begin{equation}\label{xk_sol}
x^k_n = Q_n \frac{1}{T_n + d_k {\1}_n} e_1 ||b||_2
\end{equation}

\paragraph{A Lanczos algorithm for the bilinear form}

The algorithm is derived using the Pad\'e approximation of the previous paragraph. 
First we assume that the linear system (\ref{lin_sys})
is solved to the desired accuracy using the Lanczos algorithm,
Algorithm {\ref{Lanczos_algor} and (\ref{x_sol}).
Using the orthonormality property of the Lanczos vectors and (\ref{xk_sol})
one can show that:
\begin{equation}\label{lemma}
\sum_{k=1}^m b^T \frac{c_k}{A + d_k \1} b = ||b||^2 \sum_{k=1}^m
e_1^T \frac{c_k}{T_n + d_k {\1}_n} e_1
\end{equation}
Note however that in presence of roundoff errors
the orthogonality of the Lanczos vectors is lost but the result
(\ref{lemma}) is still valid \cite{Cahill_et_al,Golub_Strakos}.
For large $m$ the partial fraction sum in the right hand side 
converges to the matrix function $f(T_n)$. Hence we get:
\begin{equation}\label{reduced_form}
{\mathcal F}(b,A) \approx
{\mathcal {\hat F}}_n(b,A) = ||b||^2 e_1^T f(T_n) e_1
\end{equation}
Note that the evaluation of the right hand side
is a much easier task than
the evaluation of the right hand side of (\ref{b_forms}).
A straightforward method is the spectral decomposition of the
symmetric and tridiagonal matrix $T_n$:
\begin{equation}\label{Omega}
T_n = Z_n \Omega_n Z_n^T
\end{equation}
where $\Omega_n \in \Real^{n\times n}$ is a diagonal matrix of
eigenvalues $\omega_1,\ldots,\omega_n$ of $T_n$ and $Z_n \in \Real^{n\times n}$
is the corresponding matrix of eigenvectors, i.e. $Z_n = [z_1,\ldots,z_n]$.
From (\ref{reduced_form}) and
(\ref{Omega}) it is easy to show that (see for example \cite{Golub_VanLoan}):
\begin{equation}\label{omega_form}
{\mathcal {\hat F}}_n(b,A) = ||b||^2 e_1^T Z_n f(\Omega_n) Z_n^T e_1
\end{equation}
where the function $f(.)$ is now evaluated at individual eigenvalues of
the tridiagonal matrix $T_n$.

The eigenvalues and eigenvectors of a symmetric and tridiagonal matrix
can be computed by the QR method with implicit shifts
\cite{eig_templates}. The method has an $O(n^3)$ complexity.
Fortunately, one can compute (\ref{omega_form}) with
only an $O(n^2)$ complexity.
Closer inspection of eq. (\ref{omega_form}) shows that besides the
eigenvalues, only the first elements of the eigenvectors are needed:
\begin{equation}\label{result_form}
{\mathcal {\hat F}}_n(b,A) = ||b||^2 \sum_{i=1}^n z_{1i}^2 f(\omega_i)
\end{equation}
It is easy to see that the QR method delivers the eigenvalues and
first elements of the eigenvectors with $O(n^2)$ complexity.

A similar formula (\ref{result_form}) is suggested
by \cite{Bai_et_al}) based on
quadrature rules and Lanczos polynomials.
The Algorithm \ref{lambda_algor} is thus another way to compute the
bilinear forms of the type (\ref{b_forms}).
\begin{algorithm}[htp]
\caption{The Lanczos algorithm for computing (\ref{b_forms}).}
\label{lambda_algor}
\begin{algorithmic}
\STATE Set $\beta_0 = 0, ~\rho_1 = 1 / ||b||_2, ~q_0 = o, ~q_1 = \rho_1 b$
\FOR{$~i = 1, \ldots$}
    \STATE $v = A q_i$
    \STATE $\alpha_i = q_i^{\dag} v$
    \STATE $v := v - q_i \alpha_i - q_{i-1} \beta_{i-1}$
    \STATE $\beta_i = ||v||_2$
    \STATE $q_{i+1} = v / \beta_i$
    \STATE $\rho_{i+1} = - (\rho_i \alpha_i + \rho_{i-1} \beta_{i-1}) / \beta_i$
    \IF{$1 / |\rho_{i+1}| < \epsilon$}
       \STATE $n = i$
       \STATE stop
    \ENDIF
\ENDFOR
\STATE Set $~(T_n)_{i,i} = \alpha_i, ~(T_n)_{i+1,i} = (T_n)_{i,i+1} = \beta_i$,
       otherwise $~(T_n)_{i,j} = 0$
\STATE Compute $\omega_i$ and $z_{1i}$ by the QL method
\STATE Evaluate (\ref{b_forms}) using (\ref{result_form})
\end{algorithmic}
\end{algorithm}

The Lanczos algorithm alone
has an $O(nN)$ complexity, whereas Algorithm \ref{lambda_algor} has
a greater complexity: $O(nN)+O(n^2)$. For typical applications
in lattice QCD the $O(n/N)$ additional relative overhead is small and therefore
Algorithm \ref{lambda_algor} is the recommended algorithm to compute the
bilinear form (\ref{b_forms}).

We stop the iteration when
the underlying liner system is solved to the desired accuracy.
However, this may be too demanding since the prime interest here
is the computation of the bilinear form (\ref{b_forms}). Therefore,
a better stopping criterion is to monitor the convergence of the
bilinear form as proposed in \cite{Bai_et_al}.
\begin{figure}
\vspace{4cm}
\epsfxsize=6cm
\hspace{3.5cm} \epsffile[200 400 480 450]{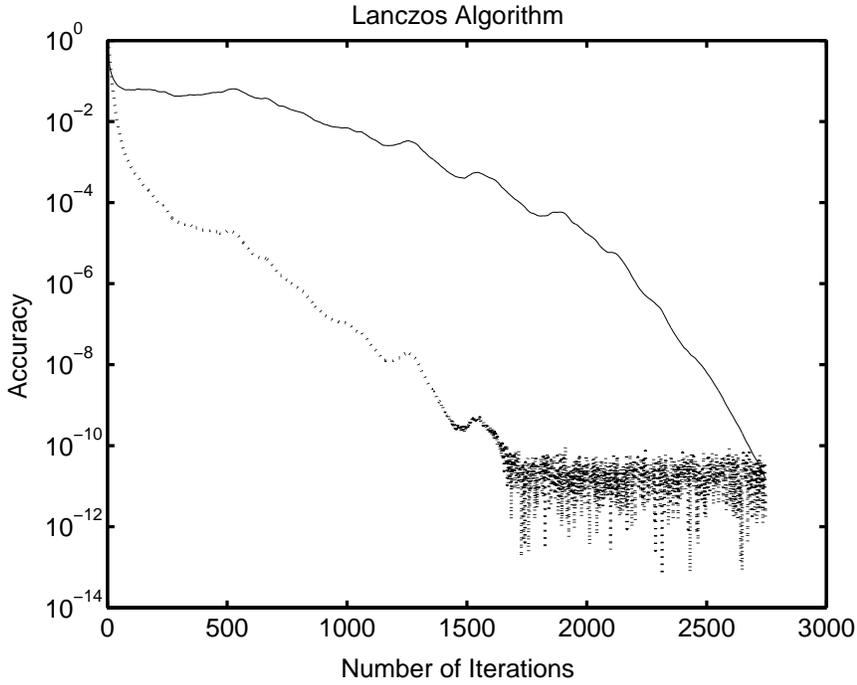}
\vspace{5cm}
\caption{Normalized recursive residual (solid line) and
relative differences of (\ref{result_form}) (dotted line)
produced by Algorithm \ref{lambda_algor}.}
\end{figure}

To illustrate this situation we give an example from a $12^3\times24$
lattice with $\mu = 0.2$, bare quark mass $m_q = -0.869$ and a $SU(3)$
gauge field background at bare gauge coupling $\beta = 5.9$.
We compute the bilinear form (\ref{b_forms}) for:
\begin{equation}
f(s) = \log \tanh \sqrt{s}, ~~~~~s \in \Real_+
\end{equation}
and $A = H_W(m_q,U)^2$, $b \in \R^N$ and
$b-$elements are chosen randomly from the set $\{+1,-1\}$.

In Fig. 1 are shown the normalized recursive residuals
$\norm{b - Ax_i}_2/\norm{b}_2$, $i=1,\ldots,n$
and relative differences of (\ref{result_form})
between two successive Lanczos steps.
The figure illustrates clearly the different regimes of convergence
for the linear system and the bilinear form.
The relative differences of the bilinear form
converge faster than the computed recursive residual.
This example indicates that a stopping criterion based on the
solution of the linear system may indeed be strong and demanding.
Therefore, the recommended stopping criteria would be to monitor the
relative differences of the bilinear form
but less frequently than proposed by \cite{Bai_et_al}. More investigations
are needed to settle this issue.
Note also the roundoff effects (see Fig. 1) in the convergence of the
bilinear form which are a manifestation of the finite precision
of the machine arithmetic.

\section{The link between overlap and domain wall fermions}

Wilson regularization of quarks violates chiral symmetry even
for massless quarks. This is a serious problem if we would like to
compute mass spectrum with light sea quarks. The improvement of the
discretization helps to reduce chiral symmetry breaking terms for a
Wilson fermion. However, one must go close to the continuum limit
in order to benefit from the improvement programme. This is
not affordable with the present computing power.

The idea of \cite{Ka92} opened the door for chiral symmetry realization
at finite lattice spacing. \cite{NaNe93} proposed overlap fermions and
\cite{FuSha95} domain wall fermions as a theory of
chiral fermions on the lattice. The overlap operator is defined by \cite{Ne98}:
\begin{equation}\label{overlap_op}
D(m_q,U) = \frac{1+m_q}{2}\1 + \frac{1-m_q}{2} \gamma_5 \text{sign}[H_W(M,U)]
\end{equation}
with $M \in (-2,0)$ which is called the domain wall height, which substitutes
the original bare quark mass of Wilson fermions. From now on we suppress the
dependence on $M,m_q$ and $U$ of lattice operators for the ease of notations.

Domain wall fermions are lattice fermions in 5-dimensional Euclidean space
time similar to the 4-dimensional formulation of Wilson fermions but with
special boundary conditions along the fifth dimension. The 5-dimensional
domain wall operator can be given by the $L_5\times L_5$ blocked matrix:
\begin{equation*}
\mathcal M =
\begin{pmatrix}
a_5D_W-\1              & P_+          &         & -m_qP_-      \\
P_-                    & a_5D_W-\1    & \ddots  &              \\
                       & \ddots       & \ddots  & P_+          \\
-m_qP_+                &              & P_-     & a_5D_W-\1    \\
\end{pmatrix}
,~~~P_{\pm} = \frac{\1_4 + \gamma_5}{2}
\end{equation*}
where the blocks are matrices defined on the 4-dimensional lattices and
$P_{\pm}$ are $4\times 4$ chiral projection operators. Their presence
in the blocks with dimensions $N\times N$ should be understood as the direct
product with the colour and lattice coordinate spaces.
$a_5$ is the lattice spacing along the fifth dimension.

These two apparently different formulations of chiral fermions are in fact
closely related to each other \cite{Borici_99}.
To see this we must calculate for the domain
wall fermions the low energy effective fermion matrix in four dimensions.
This can be done by calculating the transfer matrix $T$ along the
fifth dimension. Multiplying $\mathcal M$ from the right by the permutation matrix:
\begin{equation*}
\begin{pmatrix} P_+ & P_- &        &     \\
                    & P_+ & \ddots &     \\
                    &     & \ddots & P_- \\
                P_- &     &        & P_+ \\
\end{pmatrix}
\end{equation*}
we obtain:
\begin{equation*}
\gamma_5
\begin{pmatrix}
(a_5H_WP_+ - \1)(P_+ - m_qP_-) & a_5H_WP_- + \1 &        &              \\
                               & a_5H_WP_+ - \1 & \ddots &              \\
                               &                & \ddots & a_5H_WP_- + \1 \\
(a_5H_WP_- + \1)(P_- - m_qP_+) &                &        & a_5H_WP_+ - \1 \\
\end{pmatrix}
\end{equation*}
Further, multiplying this result from the left by the inverse
of the diagonal matrix:
\begin{equation*}
\begin{pmatrix}
a_5H_WP_+ - \1 &                &        &                \\
               & a_5H_WP_+ - \1 &        &                \\
               &                & \ddots &                \\
               &                &        & a_5H_WP_+ - \1 \\
\end{pmatrix}
\end{equation*}
we get:
\begin{equation}
\mathcal{T}(m) :=
\begin{pmatrix} P_+ - m_qP_-     & -T &        &          \\
                                 & \1 & \ddots &          \\
                                 &    & \ddots & -T       \\
                -T(P_- - m_qP_+) &    &        & \1       \\
\end{pmatrix}
\end{equation}
with the transfer matrix $T$ defined by:
\begin{equation*}
T = \frac{\1}{\1 - a_5H_WP_+} (\1 + a_5H_WP_-)
\end{equation*}
By requiring the transfer matrix being in the form:
\begin{equation*}
T = \frac{\1 + a_5\mathcal{H_W}}
         {\1 - a_5\mathcal{H_W}}
\end{equation*}
it is easy to see that \cite{Borici_99}:
\begin{equation}\label{dwf_HW}
\mathcal{H_W} = H_W \frac{1}{2 - a_5D_W}
\end{equation}
Finally to derive the four dimensional Dirac operator one has to
compute the determinant of the effective fermion theory in four
dimensions:
\begin{equation*}
\det D^{(L_5)} = \frac{\det\mathcal{T}(m_q)}{\det\mathcal{T}(1)}
\end{equation*}
where the subtraction in the denominator corresponds to a 5-dimensional
theory with anti-periodic boundary conditions along the fifth dimension.
It is easy to show that the determinant of the $L_5\times L_5$
block matrix $\mathcal{T}(m_q)$ is given by:
\begin{equation*}
\det\mathcal{T}(m_q) = \det[(P_+ - m_qP_-) - T^{L_5} (P_- - m_qP_+)]
\end{equation*}
or
\begin{equation*}
\det\mathcal{T}(m_q) =
 \det[  \frac{1 + m_q}{2} \gamma_5 (\1 + T^{L_5})
      + \frac{1 - m_q}{2}          (\1 - T^{L_5})]
\end{equation*}
which gives:
\begin{equation}
\label{4d_eff_dwf}
D^{(L_5)} = \frac{1 + m_q}{2} \1 + \frac{1 - m_q}{2} \gamma_5
        \frac{\1 - T^{L_5}}{\1 + T^{L_5}}
\end{equation}
In the large $L_5$ limit one gets the Neuberger operator (\ref{overlap_op})
but now the operator $H_W$ substituted with the operator
$\mathcal H_W$ (\ref{dwf_HW}). Taking the continuum limit in the fifth
dimension one gets ${\mathcal H_W} \rightarrow H_W$. This way overlap
fermions are a limiting theory of the domain wall fermions. To achieve
the chiral properties as in the case of overlap fermions
one must take the large $L_5$ limit in the domain wall formulation.

One can ask the opposite question: is it possible to formulate the
overlap in the form of the domain wall fermions? The answer is yes
and this is done using truncated
overlap fermions \cite{Borici_TOV}. The corresponding
domain wall matrix is given by:
\begin{equation*}
{\mathcal M}_{TOV} =
\begin{pmatrix}
a_5D_W-\1          & (a_5D_W+\1)P_+ &                & -m_q(a_5D_W+\1)P_- \\
(a_5D_W+\1)P_-     & a_5D_W-\1      & \ddots         &                    \\
                   & \ddots         & \ddots         & (a_5D_W+\1)P_+     \\
-m_q(a_5D_W+\1)P_+ &                & (a_5D_W+\1)P_- & a_5D_W-\1          \\
\end{pmatrix}
\end{equation*}
The transfer matrix of truncated overlap fermions is calculated using the
same steps as above. One gets:
\begin{equation*}
T_{TOV} = \frac{\1 + a_5H_W}
               {\1 - a_5H_W}
\end{equation*}
The 4-dimensional Dirac operator has the same form as the corresponding
operator of the domain wall fermion (\ref{4d_eff_dwf}) where $T$ is substituted
with $T_{TOV}$ (or $\mathcal H_W$ with $H_W$). Therefore the overlap Dirac
operator (\ref{overlap_op}) is recovered in the large $L_5$ limit.

\section{A two-level algorithm for overlap inversion}

In this section we review the two-level algorithm of \cite{Borici_MG}.
The basic stucture of the algorithm is that of a preconditioned Jacobi:
\begin{equation}
x^{i+1} = x^i + S_n^{-1} (b - D x^i), ~~~i=0,1,\ldots
\end{equation}
where $S_n$ is the preconditioner of the overlap operator $D$ given by:
\begin{equation}
S_n = \frac{1+m_q}{2} \1 + \frac{1-m_q}{2} \gamma_5 H_W
\sum_{k=1}^n \frac{a_k}{H_W^2 + b_k \1}
\end{equation}
where $a_k,b_k \in \Real, k=1,\ldots,n$ are coefficients that can be optimized
to give the best rational
approximation of the $sign(H_W)$ with the least number of terms
in the right hand side. For example one can use the optimal rational
approximation coefficients of Zolotarev \cite{Zo1877,PePo87}.
For the rational approximation of $sign(H_W)$ one has:
\begin{equation}
D = \lim_{n \rightarrow \infty} S_n
\end{equation}

In order to compute efficiently the inverse of the preconditioner we go back
to the 5-dimensional formulation of the overlap operator (see the previous
section as well) which can be written as a matrix
in terms of 4-dimensional block matrices:
\begin{equation}
{\cal H} =
\begin{pmatrix}
\frac{1+m_q}{2}\1 &\frac{1-m_q}{2}\gamma_5H_W& \cdots &\frac{1-m_q}{2}\gamma_5H_W\\
-a_1\1            & H_W^2 + b_1\1            &        &                          \\
\vdots            &                          & \ddots &                          \\
-a_n\1            &                          &        & H_W^2 + b_n\1            \\
\end{pmatrix}
\end{equation}
This matrix can be also partitioned in the $2\times 2$ blocked form:
\begin{equation}
{\cal H} =
\begin{pmatrix}
H_{11} & H_{12} \\
H_{21} & H_{22} \\
\end{pmatrix}
\end{equation}
with Schur complement:
\begin{equation}
S_{11} = H_{11} - H_{12} H_{22}^{-1} H_{21}
\end{equation}
It is easy to show that the following statements hold:

\begin{prop}
i) The preconditioner $S_n$ is given by the Schur complement:
\begin{equation}
S_n = S_{11}
\end{equation}
ii) Let ${\cal H} \chi = \eta$
with $\chi = (y,\chi_1,\ldots,\chi_n)^T$ and $\eta = (r,o,\ldots,o)^T$.
Then $y$ is the solution of the linear system $S_n y = r$.
\end{prop}
Using these results and keeping $n$ fixed the algorithm of \cite{Borici_MG}
(known also as the multigrid algorithm)
has the form of a two level algorithm:
\begin{algorithm}[htp]
\caption{\hspace{.2cm} A two-level algorithm for overlap inversion}
\label{two_level_algor}
\begin{algorithmic}
\STATE Set $x^1 \in \Complex^N,
~r^1 = b, ~\text{tol} \in \Real_+, ~\text{tol}_1 \in \Real_+$
\FOR{$~i = 1, \ldots$}
    \STATE Solve approximately ~${\cal H} \chi^{i+1} = \eta^i$ such that
           $||\eta^i - {\cal H} \chi^{i+1}||_2/||r^i||_2 < \text{tol}_1$
    \STATE $x^{i+1} = x^i + y^{i+1}$
    \STATE $r^{i+1} = b - D x^{i+1}$
    \STATE Stop if $||r^{i+1}||_2/||b||_2 <$ tol
\ENDFOR
\end{algorithmic}
\end{algorithm}
This algorithm is in the form of nested iterations. One can see that the
outer loop is Jacobi iteration which contains inside two inner iterations:
the approximate solution of the 5-dimensional system and the multiplication
with the overlap operator $D$ which involves the computation of the $sign$
function. For the 5-dimensional system one can use any iterative solver
which suites the properties of $\cal H$. We have used many forms for $\cal H$
ranging from rational approximation to domain wall formulations.
For the overlap multiplication we have
used the algorithm of \cite{Borici_isqr}. Our test on a small lattice show that
the two level algorithm outperforms with an order of magnitude the brute force
conjugate gradients nested iterations \cite{Borici_MG}.

Since the inner iteration solves the problem in a 5-dimensional lattice with
finite $L_5$ and the outer iteration solves for the 4-dimensional projected
5-dimensional system with $L_5 \rightarrow \infty$, the algorithm in its nature
is a multigrid algorithm along the fifth dimension. The fact that the multigrid
works here is simply the free propagating fermions in this direction. If this
direction is gauged, the usual problems of the multigrid on a 4-dimensional
lattice reappear and the idea does not work. In fact, this algorithm with
$n$ fixed is a two grid algorithm. However, since it does not involve the classical
prolongations and contractions it can be better described as a two-level algorithm.
\footnote{I thank Andreas Frommer for discussions on this algorithm.}

\printindex
\end{document}